\documentclass[letterpaper, 10 pt, conference]{IEEEtran}  

\IEEEoverridecommandlockouts
\usepackage{amsmath,amssymb,amsfonts}
\usepackage{algorithmic}
\usepackage{graphicx}
\usepackage{textcomp}
\usepackage{xcolor}
\usepackage{todonotes}
\usepackage{comment}
\usepackage{mathrsfs}
\usepackage{xspace}
\usepackage{subfigure}
\usepackage{hyperref}
\usepackage{fancyhdr}    
\usepackage{colortbl}
\definecolor{lightred}{rgb}{1.0, 0.6, 0.6}
\definecolor{lightgreen}{rgb}{0.6, 1.0, 0.6}
\definecolor{lightyellow}{rgb}{1.0, 1.0, 0.6}
\usepackage[backend=biber,style=ieee]{biblatex}
\usepackage[font=scriptsize	,labelfont=bf]{caption}

\usepackage[explicit,noindentafter]{titlesec}
\titlespacing\section{2pt}{4pt plus 2pt minus 2pt}{4pt plus 2pt minus 2pt}
\titlespacing\subsection{1pt}{4pt plus 2pt minus 2pt}{4pt plus 2pt minus 2pt}
\titlespacing\subsubsection{1pt}{4pt plus 2pt minus 2pt}{4pt plus 2pt minus 2pt}

\newcommand{\name}{PhaseMO\xspace}
\def\BibTeX{{\rm B\kern-.05em{\sc i\kern-.025em b}\kern-.08em
    T\kern-.1667em\lower.7ex\hbox{E}\kern-.125emX}}
\begin{document}
\def\bibfont{\scriptsize}

\title{\name: A Universal Massive MIMO Architecture for Sustainable NextG\vspace{-10pt}
}

\author{
\IEEEauthorblockN{Adel Heidari, Agrim Gupta, Ish Kumar Jain, and Dinesh Bharadia}
\IEEEauthorblockA{
University of California, San Diego, USA \\
\{adheidari, agg003, ikjain, dbharadia\}@ucsd.edu}
}


\maketitle
\begin{abstract}
The rapid proliferation of devices and increasing data traffic in cellular networks necessitate advanced solutions to meet these escalating demands. Massive MIMO (Multiple Input Multiple Output) technology offers a promising approach, significantly enhancing throughput, coverage, and spatial multiplexing. Despite its advantages, Massive MIMO systems often lack flexible software controls over hardware, limiting their ability to optimize operational expenditure (OpEx) by reducing power consumption while maintaining performance. Current software-controlled methods, such as antenna muting combined with digital beamforming and hybrid beamforming, have notable limitations. Antenna muting struggles to maintain throughput and coverage, while hybrid beamforming faces hardware constraints that restrict scalability and future-proofing. This work presents \name, a versatile approach that adapts to varying network loads. \name effectively reduces power consumption in low-load scenarios without sacrificing coverage and overcomes the hardware limitations of hybrid beamforming, offering a scalable and future-proof solution. We will show that \name can achieve up to 30\% improvement in energy efficiency while avoiding about 10\% coverage reduction and a 5dB increase in UE transmit power.

\end{abstract}

\begin{IEEEkeywords}
Universal Beamforming, massive MIMO, Sustainable NextG, Spatial Multiplexing, Digital Beamforming, Hybrid Beamforming
\end{IEEEkeywords}
\section{Introduction}\label{sec:introduction}

With every cellular generation, the number of antennas increases, since more antennas allow access to spatial degrees of freedom. This provides benefits like increased coverage, higher throughput, and spatial multiplexing to help scale to a large number of users and exponential growth in mobile networks.
At present, the most deployed multi-antenna technology is Massive MIMO, which utilizes a `massive' number of antennas, that can be as high as $64-128$, to provide increased coverage $>1$ km, net throughputs over $1$ Gbps, and the ability to multiplex $8-16$ users in the spatial domain \cite{samsungMIMO}.

For considering the ease of deployment, most often these performance metrics (throughput, coverage) are reported when the Massive MIMO array is being fully utilized, and considered as the peak performance.
As cellular networks mature and evolve into the next generation, softwarized control over the radio hardware has emerged as an important theme \cite{yang2013openran}. Softwarized control provides greater flexibility over the hardware \cite{costa2013radio}, and reduction of operational expenditure (OpEx) by tuning down the power consumption when network conditions don't require peak performance \cite{rost2015benefits}.
In the context of Massive MIMO, such softwarized control aims to judiciously use the massive spatial degrees of freedom to optimize for the existing network load conditions.
For example, a Massive MIMO base station can reduce the number of spatially multiplexed layers under low load conditions, like night-time, and, hence save power.
Further, this performance toning down should be flexible, and if needed, the underlying hardware needs to start working at the peak performance once the network load increases.

To make massive MIMO adapt to network load conditions, there are mainly two broad approaches studied in the literature: (1) Antenna muting-assisted Digital Beamformers~\cite{amfrenger} and (2) Hybrid Beamformers~\cite{mendez2016hybrid}. 
The majority of the existing Massive MIMO deployments utilize Digital Beamforming architecture, which has a separate digital RF chain interface for each antenna. 
Antenna muting approaches consist of softwarized control atop Digital Beamformers, which turn off a certain number of RF chains when the network load is low. 
Antenna muting adjusts the number of antennas as the network load varies, to improve energy efficiency by not using more than the required number of antennas.
However, this leads to reduced user-perceived throughput, as well as increased user-equipment power, since the overall antenna gain reduces due to muting, and this has been reported across multiple companies in the latest 3GPP reports~\cite{3gpptr}.
The second solution, Hybrid Beamforming (HBF) aims to always utilize a large number of antennas while connecting them to a smaller number of RF chains via an analog network typically consisting of phase shifters.
Since HBF doesn't reduce the number of antennas, but only the number of RF chains, it doesn't have the required drastic effect on throughput and user device power.
However, HBF architectures are not flexible, and future-proof, that is, say we have an HBF that connects 64 antennas to 8 RF chains, it can not be scaled up to utilize the \textit{same hardware} for 16 spatial multiplexed layers.
That is, HBF architectures can only be designed for a particular network load, and are unable to scale up if needed in the future, which limits their real-world deployment. 

In this work, we present \name, which enables the best elements from the prior two solutions, that is, flexible reduction of power, adaptive to network load, akin to antenna muting, and as well having the ability to use the entire antenna array like the hybrid beamformer, while reducing the RF chains. 
That is, in \name, the total digital compute can be optimized using software control to reduce the total number of RF chains, while always being connected to all the antennas using the proposed analog network architecture.
Hence, \name maximally utilizes all the antennas' spatial degrees of freedom to avail the maximum beamforming gain while reducing the digital processing power demanded by RF chains. 
This allows \name to operate at higher energy efficiencies than the existing solutions without creating any adverse effects on throughput, coverage, and user device power consumption. A summary of \name's provided features in comparison to existing approaches is shown in Table \ref{table:beamforming_comparison}.

To achieve high performance with reduced hardware complexity and scalability features, \name introduces a novel MIMO architecture that combines a single RF chain utilizing a high-sampling ADC/DAC and a network of Fast Phase Shifters (FPSs) which are replacing the traditional phase shifters and can provide phase shifts at sub-nanosecond speeds (faster than 1 GHz). These FPSs which are commercially available \cite{hmc877lc3} enable the creation of $V$ flexible "virtual" RF chains within a period of a single symbol. For example, with $B=100$~MHz (symbol time = 10~ns) and $V=10$, the FPSs update phase settings every 1~ns synchronous to ADC/DAC operating at $VB$ net digital conversion rate, thus creating 10 different analog beam configurations within the net 10 ns time. This process essentially creates $V=10$ different beam signals within a symbol time hence making \name, a future-proof architecture as it allows software control to scale the number of virtual RF chains $V$, by simply increasing the ADC/DAC sampling rate by $V$ times and running the FPSs $V$ times faster, without needing any hardware upgrades.

\begin{table}[t!]
\centering

\begin{tabular}{|l|p{1.5cm}|p{01.5cm}|p{1.8cm}|}
\hline
\textbf{Beamforming} & \textbf{Data Streams}  & \textbf{Energy\newline Efficiency} & \textbf{Adaptability\newline Future-Proof}  \\ \hline
Digital               & \cellcolor{lightgreen}Multiple                          & \cellcolor{lightred}Low          & \cellcolor{lightred}No                                          \\ \hline
Digital + AM          & \cellcolor{lightgreen}Multiple                          &\cellcolor{lightyellow} Medium          & \cellcolor{lightgreen}Yes                                         \\ \hline
Hybrid                & \cellcolor{lightyellow}Multiple\newline (Restricted)      &\cellcolor{lightgreen} High        & \cellcolor{lightred}No                           \\ \hline
\textbf{\name}             & \cellcolor{lightgreen}Multiple\newline (Unrestricted)           & \cellcolor{lightgreen}High           & \cellcolor{lightgreen}Yes\newline     \\ \hline
\end{tabular}
\caption{Comparison of beamforming techniques for spatial multiplexing: \name achieves higher energy efficiency and adaptive data stream scaling compared to DBF, AM-assisted DBF, and Hybrid Beamforming.}
\label{table:beamforming_comparison}
\end{table}

In this paper, we describe the required mathematical models to show the exact process behind the construction of these $V$ RF chains, the required approximations, analog non-idealities, and their overall impact on the system performance. We show that by always utilizing the large number of antennas, \name performance matches throughput and coverage metrics of state-of-art hybrid beamformers, while capable of tuning up and down as needed.
That is when $V=1$, \name takes the form of an analog beamformer, and when $V=N$ it becomes like a digital beamformer, while any intermediary value \name emulates a hybrid beamformer. Overall, this ability to control the digital compute based on the choice of $V$ makes \name energy efficient, able to reduce power consumption while not introducing any detriments towards throughput, coverage, and user device power. The paper is organized as follows: Section \ref{sec:related} reviews related work, Section \ref{sec:background} outlines the system model, Section \ref{sec:design} details the \name architecture, Section \ref{sec:evaluation} evaluates \name's performance, and Section \ref{sec:discussion} discusses its limitations and implications.

\section{Related Work}\label{sec:related}

\begin{figure*}[t!]
    \centering
    \includegraphics[width=\linewidth]{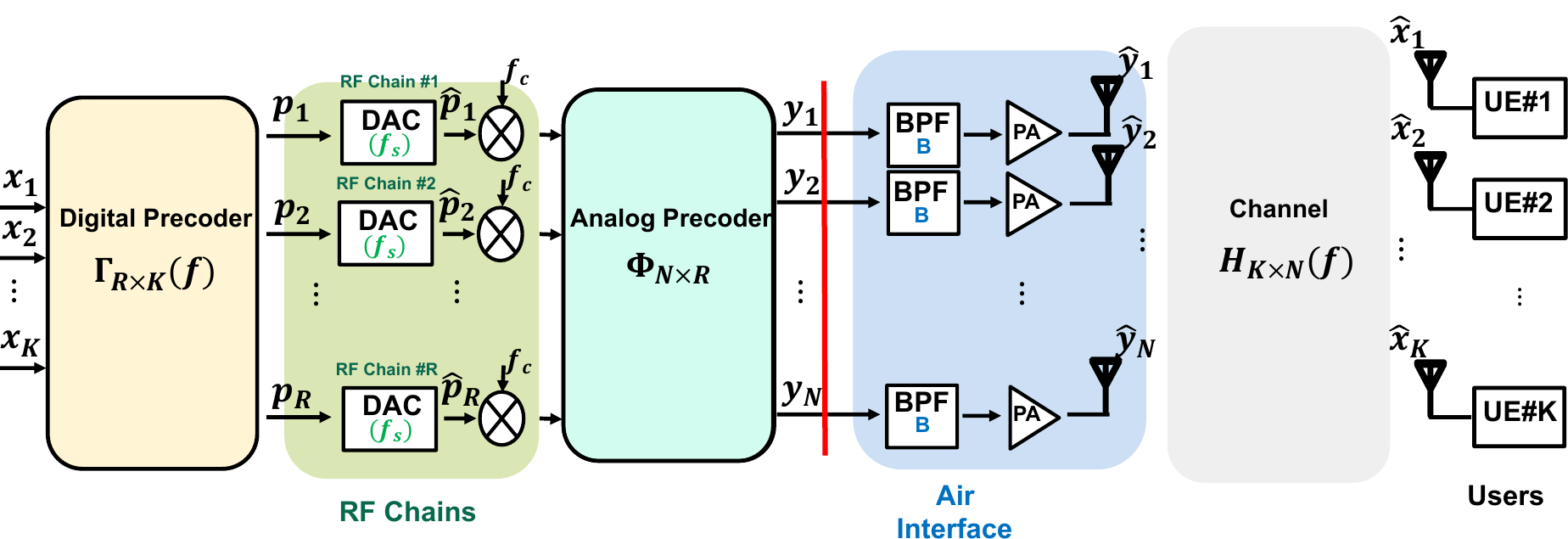}
    \caption{A generic architecture model for beamforming structures, encompassing four parts:
1) \textbf{Baseband Digital Precoder}: Digital precoding over subcarriers on users' data vector $X_{K\times 1}(f)$ with bandwidth $B$.
2) \textbf{RF Chains}: Precoded digital samples $P_{R\times 1}(f)$ pass through DACs (sampling frequency $f_s$) and upconverters to passband $f_c$, resulting in analog output signals $\hat{P}_{R\times 1}(f)$.
3) \textbf{RF Analog Beamformer}: Phase shifters perform analog precoding, mapping $R$ RF chains' analog signals to $N$ signals $Y_{N\times 1}(f)$ radiated from antennas.
4) \textbf{Air Interface Part}: Power amplifiers, antennas, and bandpass filters operate on $Y_{N\times 1}(f)$ to produce $\hat{Y}_{N\times 1}(f)$ with bandwidth $B$ centered around $f_c$.}

    \label{fig:generic_fig}
\end{figure*}

Improving the energy efficiency of wireless networks is gaining interest from both academic research \cite{han2020energy,bjornson2018energy}, as well as industry standards having work items on network energy savings \cite{3gpptr}. Researchers are actively exploring methods and techniques to reduce energy consumption by judicious use of the temporal \cite{wang2016energy,kundu2024towards}, frequency \cite{choi2023deep,azimi2021energy}, and spatial resources \cite{wesemann2023energy,lopez2022survey} available to a radio. 

In the context of Massive MIMO, the optimization of spatial resources is of primary importance. Here, antenna muting has emerged as an important method of reducing energy consumption, by optimizing the number of active antennas and hence the associated RF chains \cite{amfrenger,guruprasad2017user,akhtar2020joint}.
However, since antenna muting reduces the number of active antennas, it also causes adverse effects on throughput and coverage, as reported by multiple companies in the network energy study by 3GPP \cite{3gpptr}. 
Further, to reduce this adverse impact, the required antenna selection algorithms to determine which antennas to mute are not straightforward. 
These algorithms often involve a data-driven approach to model the network load and fine-tune the algorithms properly to ensure that antenna muting adverse effects are reduced \cite{amseedrl,amNN,amRNN}. 

In comparison, there are alternate sets of works, that utilize all the antennas, but reduce the number of RF chains instead, by using Hybrid beamforming (HBF) approaches \cite{kuhne2020bringing,mendez2016hybrid}.
Fully-connected HBF capture maximum array gain per RF chain \cite{mondal201921, mondal201825}, unlike the partially connected HBF \cite{kim201828,xie2015hekaton} counterparts, which only have a subset of antennas connected.
Hence, Fully-connected HBF increases energy efficiency by reducing the digital processing required \cite{chen2020turbo,yu2016alternating}. 
However, Fully-connected HBF is shown to have challenges in hardware implementations since they require complex analog networks with multiple splitter networks \cite{yan2019performance,barati2020energy}, to ensure all the antennas are available to all the RF chains.

In addition to HBF, there are other proposed antenna arrays, that utilize RF switches \cite{gupta2023greenmo, bogdan2020time,gonzalez2020wideband} to multiplex multiple antennas via a single RF chain using time domain codes.
Most notable of these is GreenMO \cite{gupta2023greenmo}, which implements such switched arrays for wideband OFDM waveforms and shows the feasibility of multiple antennas sharing a single RF chain to achieve energy efficiency.
However, commercially available RF switches can only reach switching speeds of $\sim 10$ns, which limits the number of antennas that can share the same RF chain. This limits the scalability of these ideas.
Further, RF switches only allow for antenna-selection-based beamforming gains, which fail to capture the maximum beamforming gains possible from co-phased combining across antennas.

In this paper, we show how using commercially available Fast Phase Shifters (FPS) \cite{hmc877lc3} can effectively lead to both, faster multiplexing to increase the number of multiplexed antennas, as well as efficient co-phased combining across antennas to achieve full beamforming gain. They typically have voltage-controlled circuits that allow for high-speed phase changes in the order of 1 ns (symbol-level) at the cost of a little bit higher power consumption, quite faster than digital-controlled phase shifters \cite{planar_phase_shifter} which usually work at the speed of slot time.

\section{System Model}\label{sec:background}
To analyze various beamforming architectures and their performance, we use a generic model encompassing all existing structures, as shown in Fig. \ref{fig:generic_fig}. We illustrate how hardware architectural differences affect the mathematical model and performance. Using this model, we derive the signal expressions emitted from the antennas after digital and analog precoding. DAC non-idealities are excluded at this stage, assuming they are mitigated by the BPF in the air interface.

Consider a downlink scenario with \(N\) antennas and \(K\) users. The emitted signal is:

\begin{equation}
  \hat{Y}_{N \times 1}(f) = \Phi_{N \times R} \Gamma_{R \times K}(f) X_{K \times 1}(f),
   \label{eq1}
\end{equation}

where \(\Phi_{N \times R}\) is the analog precoding matrix for \(N\) antennas and \(R\) physical RF chains, \(\Gamma_{R \times K}(f)\) is the digital precoding matrix, and \(X_{K \times 1}\) is the data for \(K\) user streams.

\begin{itemize}
    \item \textbf{Baseband Digital Precoder}: The precoding matrix \(\Gamma_{R \times K}(f)\) applies baseband digital precoding across subcarriers to the users' data vector \(X_{K \times 1}(f)\), which has a bandwidth \(B\). This process generates \(R\) precoded signals, \(P_{R \times 1} = \Gamma_{R \times K}(f) X_{K \times 1}(f)\) each corresponding to an RF chain.

    \item \textbf{RF Chains}: The precoded digital samples are processed through RF chains, each comprising a digital-to-analog converter (DAC) with a sampling frequency \(f_s\) and an upconverter that shifts the baseband signal to the passband at \(f_c\). The resulting analog signals, \(\hat{P}_{R \times 1}= A(\Gamma_{R \times K}(f) X_{K \times 1}(f))\), include DAC non-idealities, such as sideband images near harmonics of the sampling frequency which is assumed this effect is modeled with the function $A$.

    In other words, although the DAC's effect varies depending on the technology used in its operation, it is generally observed that the DAC output creates sideband spectrums of the main signal at sampling frequency products, with a sinc roll-off factor. For instance, if the DAC operates at a sampling frequency \(f_s\) on a signal with bandwidth \(B\), it will generate sidebands at multiples of \(f_s\), each sideband having a bandwidth of \(B\).

    \item \textbf{RF Analog Beamformer}: This involves a network of phase shifters that perform analog precoding on the passband signal. The analog precoding matrix, \(\Phi_{N \times R}\), consisting of unity magnitude components with varying phases, maps the \(R\) RF chains' analog signals to \(N\) signals, denoted as \(Y_{N \times 1}= \Phi_{N \times R} A(\Gamma_{R \times K}(f) X_{K \times 1}(f)) \), which are then radiated by the antennas.

    \item \textbf{Air Interface Part}: This part comprises power amplifiers (PAs), antennas, and bandpass filters (BPFs). The BPF, placed before the PA, removes non-idealities from the DAC, upconverter, and other sources. With a bandwidth of \(B\) centered at \(f_c\), it processes \(Y_{N \times 1}\) to produce \(\hat{Y}_{N \times 1}\). Using BPF, we can assume the non-idealities sourced from the DAC shown with function $A$ can be eliminated. Therefore, the emitted signal can be written as \ref{eq1}. It's worth noting that the setup for the air interface part is consistent for all the beamforming techniques.
\end{itemize}

Finally, we can write the received signals on the user side by considering the channel effect on the emitted signals.

\begin{equation}
    \hat{X}_{K\times1}(f) = H_{K\times N}(f) \Phi_{N\times R} \Gamma_{R\times K}(f) X_{K\times 1}(f)
    \label{eq2}
\end{equation}

where \(H_{K\times N}(f)\) is the wireless channel between \(N\) antennas and \(K\) users. In the following subsections, we will discuss how different parts differ with respect to different beamforming techniques and how these differences affect the mathematical model and performance for each of them.

\subsubsection{Digital beamforming} 
In a digital beamforming architecture, all precoding is performed on the digital symbols, eliminating the need for an analog precoder. As a result, the number of RF chains must equal the number of antennas, implying \(R = N\). Consequently, the analog precoding matrix \(\Phi_{N \times R}\) reduces to the identity matrix \(I_{N \times N}\), and the digital precoding matrix \(\Gamma_{R \times K}\) becomes \(\Gamma_{N \times K}\). Therefore, the mathematical expression for the signals emitted from the antennas in digital beamforming can be rewritten using Eq. \ref{eq1} as \( \hat{Y}_{N \times 1}(f) = \Gamma_{N \times K}(f) X_{K \times 1}(f)\).

\subsubsection{Analog beamforming} 
Analog beamformers, which are structurally different from digital beamformers, do not utilize any digital precoder. Instead, a single RF chain is employed to convert digital symbols into an analog signal, limiting the architecture to support only one user's data transmission at a time (\(R = K = 1\), \(X_{K \times 1}(f) = X_{1 \times 1}(f)\), and \(\Gamma_{1 \times 1} = 1\)) which significantly reduces the system's throughput. In the analog domain, a network of phase shifters is used for precoding. This network comprises as many phase shifters as there are antennas (\(N\)) in the architecture, resulting in the analog precoder matrix being an \(N \times 1\) vector. However, the precoding performance is limited since the analog precoding matrix, \(\Phi_{N \times 1}\), is not frequency-dependent, reducing its effectiveness for wideband systems. Thus, the analog beamformer expression can be rewritten using Eq.~\eqref{eq1} as \(\hat{Y}_{N \times 1}(f) = \Phi_{N \times 1} X_{1 \times 1}(f)\).

\subsubsection{Hybrid beamforming} 
Hybrid beamforming combines a digital precoder and an analog beamformer, and its structure is accurately represented by the generic model in Fig.~\ref{fig:generic_fig}. This architecture features \(R\) RF chains that convert digitally precoded symbols, \(\Gamma_{R \times K}(f)\), into analog signals, which are then beamformed via a phase shifter network and radiated from \(N\) antennas. The analog network can be fully connected, in which all \(R\) RF chains connect to all \(N\) antennas, or partially connected, in which each RF chain connects to a subset of antennas. Importantly, the maximum number of users supported is limited by the number of RF chains, \(R\), and the analog precoder \(\Phi_{N \times R}\), not being frequency-dependent, reduces performance in wideband systems. The mathematical representation of hybrid beamforming is given as \(\hat{Y}_{N\times1}(f) = \Phi_{N\times R} \Gamma_{R\times K}(f) X_{K\times 1}(f)\), where \(\Phi_{N \times R}\) has sparse non-zero elements in partially connected architectures.

Next, we demonstrate the \name architecture, showcasing it as a viable alternative to all beamforming techniques with hardware complexity comparable to analog beamforming.

\section{\name Design}\label{sec:design}
\name designs a new beamforming architecture that enables high coverage area and high throughput with a single RF chain that can adapt itself to network throughput demand. In this section, we first discuss how \name mimics the RF analog beamformer of traditional hybrid precoding through a simpler novel analog architecture using only a single RF chain to create \(N\) antenna signals in the time domain. Also, we will explain how \name gains softwarized control over the radio hardware to achieve flexible throughput beating hybrid beamforming hardware limitations. Next, we will present the combining method \name used to convert \(V\) digitally precoded signals of virtual RF chains into a single physical RF chain and demonstrate how it can be represented in the frequency domain for further explanation of the mathematical representation of the design. Finally, we will demonstrate the mathematical foundation of end-to-end communication via \name and illustrate how it has the potential to support different throughput ranges from digital beamforming to analog beamforming.

\begin{figure} [!t]
\centering
    \subfigure[\textbf{Traditional Fully-Connected Hybrid-MIMO Architecture}]{
        \includegraphics[width=0.45\textwidth]{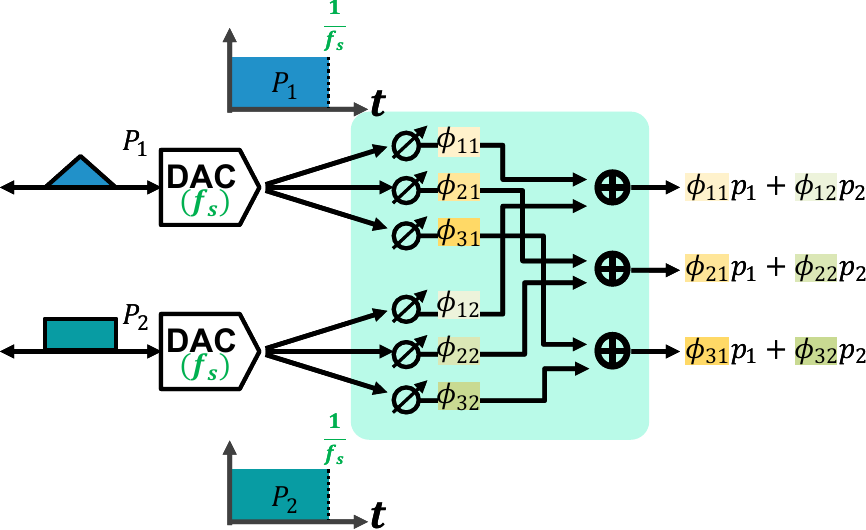}
        \label{fig:50}
      }\hfill
     \subfigure[\textbf{\name MIMO Architecture}]{
         \centering
         \includegraphics[width=.45\textwidth]{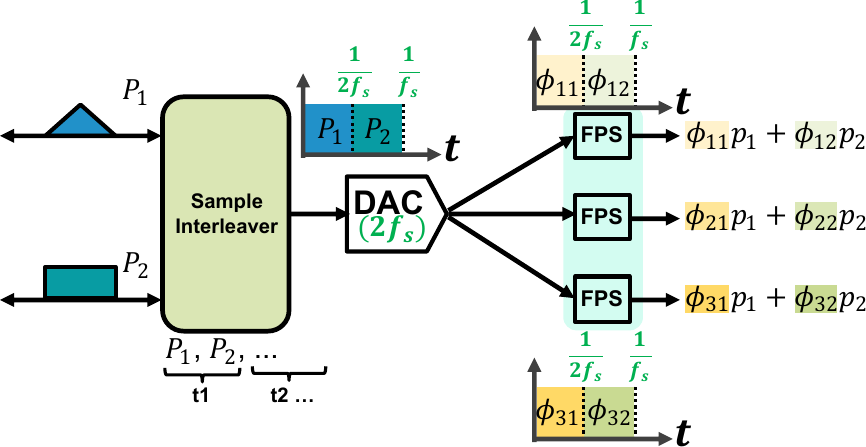}
         \label{fig:thput}
     }
     \hfill
    \caption{Traditional and \name\ beamforming architectures: a) Traditional beamforming architecture utilizes multiple RF chains, each corresponding to a precoded signal, followed by time-invariant phase shifters at each antenna. b) \name\ architecture combines all precoded signals into one via a digital interleaver, then passes it through a DAC operating at a higher frequency. Fast phase shifters at each antenna beamform the combined analog signal}
    \label{fig:phaseMO_explanation}
    \vspace{-0.2cm}
 \end{figure}

\subsection{\textbf{Beating hybrid beamforming's hardwarized flexibility}}

To practically realize how \name can generate multiple signals right after the single RF chain, and how it mimics the analog beamformer part of a traditional hybrid precoder, we consider Fig. \ref{fig:phaseMO_explanation}. As shown in this figure, a case of three antennas with two digitally precoded signals is considered. \ref{fig:phaseMO_explanation}.a) shows how traditional beamforming architecture utilizes two RF chains each of which with \(f_s\) sampling frequency and a \(3\time 2\) network of phase-shifters (PSs) to create three signals for radiation from the antennas. In this figure, we can see each RF chain is connected to all three antennas, and the signals that get radiated from the antennas are the summation of two phase-shifted RF chain signals which are produced via \(3\times 2=6\) time-invariant PSs. 

On the other hand, as shown in Fig. \ref{fig:phaseMO_explanation}.b), \name uses only a single RF chain with two times more sampling frequency to convert digitally interleaved signals into a single analog signal which goes into \(3\) FPSs each of which has individual phase control and works at DAC sampling frequency of \(2f_s\). In other words, the first FPS keeps changing its phase within $\phi_{11}$ and $ \phi_{12}$ periodically and holds each phase for \(\frac{1}{2f_s}\). As shown in Fig. \ref{fig:phaseMO_explanation}, by configuring the FPSs to shift the phase of signals in a specific pattern, we can exactly create the same signal produced in Fig. \ref{fig:phaseMO_explanation}.a). 

Furthermore, we can see if we had more digitally precoded signals (demand for more throughput), we could easily combine them to pass through the single physical RF chain with an increased sampling frequency proportionate to the number of precoded signals (Fig. \ref{fig:phaseMO_explanation}.b). However, we need to increase the number of physical RF chains in traditional hybrid beamforming architecture to be able to achieve more degrees of freedom on digital beamforming or multiplexing more beams.

\begin{figure*}[h]
    \centering
    \includegraphics[width=\linewidth]{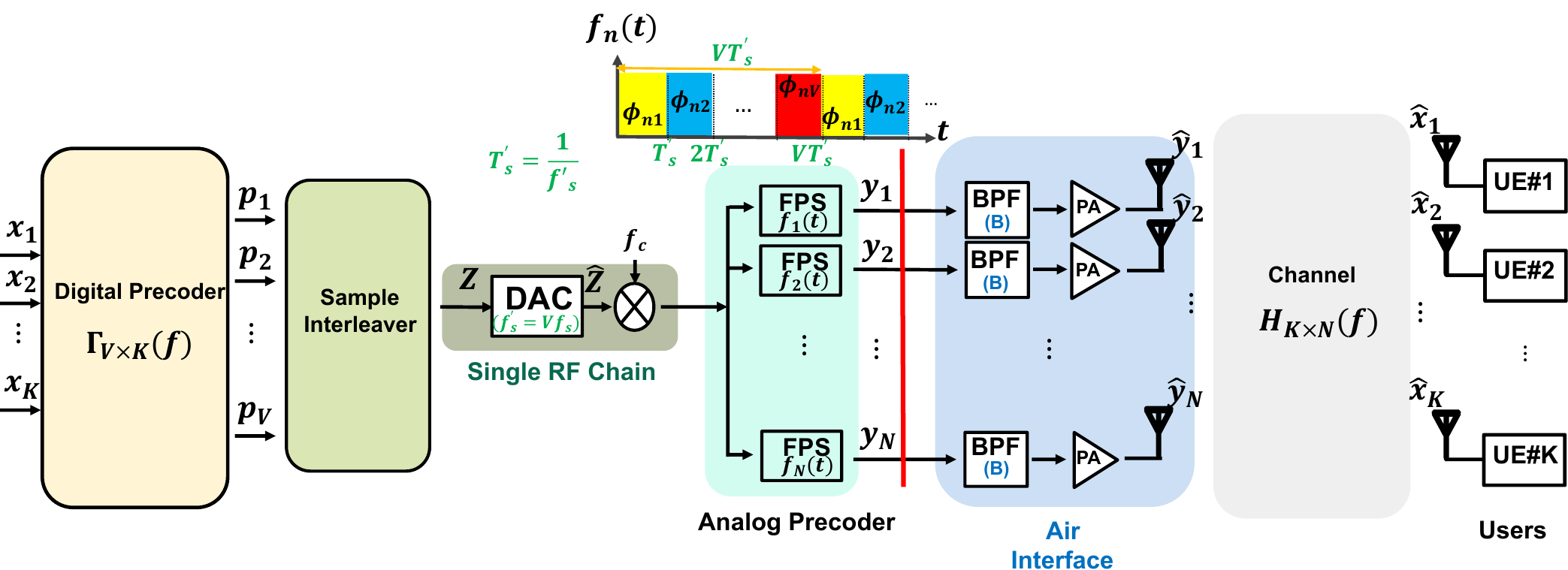}
    \caption{Architecture of the PhaseMO system, including:
1) \textbf{Baseband Digital Precoder}: The $V \times K$ matrix $\Gamma_{V \times K}(f)$ performs digital precoding on users' data vector $X_{K \times 1}$ with bandwidth $B$;
2) \textbf{Interleaver}: Organizes the $V$ digitally precoded symbols into a sequence for processing;
3) \textbf{RF Chain}: Interleaved digital samples $Z$ pass through a single RF chain with a DAC (sampling frequency $f_s$) and upconverter to passband $f_c$, resulting in analog signals $\hat{Z}$;
4) \textbf{RF Analog Beamformer}: One fast phase shifter (FPS) per antenna generates a time-variant signal $f_n(t)$, forming the matrix $\Phi_{N \times K}$, mapping the RF chain’s analog signals to $N$ output signals $Y_{N \times 1}$;
5) \textbf{Air Interface Part}: Power amplifiers (PAs), antennas, and a bandpass filter clean the spectrum, processing $Y_{N \times 1}$ to produce $\hat{Y}_{N \times 1}$.}
    \label{fig:PhaseMO_fig}
\end{figure*}

\subsection{\textbf{\name's ability to mimic any traditional beamformer}}
Now, we can generalize the toy example discussed in Fig. \ref{fig:phaseMO_explanation}.b) to unify the mathematical representation of the \name architecture, and show how \name can be configured to achieve any of the traditional beamforming schemes, explained in section \ref{sec:background} and shown with the mathematical representation of Eq. \ref{eq1}. A major assumption here is that the DAC's non-idealities combined with FPS spectrum spreading effects can be eliminated using the bandpass filter after the FPS per-antenna, which will be detailed once we present the mathematical generalization.

As shown in Fig. \ref{fig:PhaseMO_fig}, we consider \(N\) antennas, \(K\) users, and \(V\) virtual RF chains. If we denote \(V<=N\) as the number of digitally precoded symbols, we can combine all \(V\) precoded signals into a single stream using a sample interleaver. So, we can write the time domain interpretation of the interleaved signal as \(z[n]= [p_1[1], p_2[1], \ldots, p_V[1], p_1[2], p_2[2], \ldots, p_V[2], \ldots]\) which \(p_v[m]\) denotes the \(m\)-th sample of the \(v\)-th digitally precoded time-domain vector (virtual RF chain). In other words, all the $V$ vectors are upsampled by $V$ and each of them is delayed by $0,1,2,.., V$ symbols respectively. Once this signal goes to the DAC, each element in the vector will occupy $\frac{1}{Vf_s}$ time duration, so a single element delay in the vector will act as $\frac{1}{Vf_s}$ time delay. Since we want to analyze the signal's spectrum, we must write down the frequency domain representation of sample interleaver output in the analog domain ignoring the DAC's impacts as follows:
\begin{equation}
    Z(f) = P_{1}(f) + e^{-\frac{j 2 \pi f}{Vf_s}} P_{2}(f) + \cdots + e^{-\frac{j (V-1) 2 \pi f}{Vf_s}} P_{V}(f)
    \label{eq: dac output frequency}
\end{equation}

Which \(e^{-\frac{j v 2 \pi f}{Vf_s}}\) represents the frequency-domain phase shift due to the time-domain delay for the vector of precoded symbols \(v\) -th. Notably, the effect of
V-times upsampling is compensated by increasing the DAC sampling frequency by
V-times.

In the next step, we consider the effect of FPS on the DAC output. As mentioned previously, each FPS toggles periodically between $V$ phases each of which with \(\frac{1}{Vf_s}\) duration, for instance, FPS located at nth antenna creates $\Phi_{n1}, \Phi_{n2}, ...,$ and $\Phi_{nV}$ with period of \(\frac{1}{f_s}\). Therefore, the radiated signal from the nth antenna can be determined as follows, which is the multiplication of constant periodic phases and DAC's output:
 
\begin{multline}
    \hat{Y}_n(f) = \sum_{v=0}^{V-1} e^{j\Phi_{nv}} \left(P_{1}(f) + e^{-\frac{j 2 \pi f}{Vf_s}} P_{2}(f) \right. \\
    \left. + \cdots + e^{-\frac{j (V-1) 2 \pi f}{Vf_s}} P_{V}(f)\right)
    \label{eq: radiated signal wo nonideal}
\end{multline}

By considering \(\hat{P}_v(f)=P_v(f) e^{\frac{j v 2 \pi f}{Vf_s}}\), we can express the summation in the matrix form of \(\hat{Y}_{N\times 1}(f) = \Phi_{N\times V} \hat{P}_{V \times 1}(f)\), which \(\Phi_{N\times V}\) represents the phase matrix represented by \(N\) FPSs. Furthermore, we can also include a digital precoding matrix denoted by \(\Gamma_{V\times K}(f)\) in this part to clarify the final radiated signal (\(\hat{P}_{V\times1}=\Gamma_{V\times K}(f)X_{K\times 1}(f)\)):
\begin{equation}
    \hat{Y}_{N\times 1}(f) = \Phi_{N\times V} \Gamma_{V\times K}(f) \mathbf{X}_{K \times 1}(f)
    \label{eq11}
\end{equation}

We now aim to further explain the equation derived for the signal emitted from the antennas (Eq.~\ref{eq11}) and discuss how PhaseMO can relate to the other beamforming techniques mentioned earlier. 
\begin{itemize}
    \item For \textbf{V=N}, the final equation can be written as \(\hat{Y}_{N\times 1}(f) = \Phi_{N\times N} \Gamma_{N\times K}(f) \mathbf{X}_{K \times 1}(f)\). This equation is exactly similar to what we already derived for a digital beamformer if we consider $\Phi_{N\times N}$ as an identity matrix which can be achieved using 
    \item For \textbf{V=1}, the final equation can be written as \(\hat{Y}_{N\times 1}(f) = \Phi_{N\times 1} \mathbf{X}_{1 \times 1}(f)\). We can observe that this equation is in the same form as that for an analog beamformer, which uses just one RF chain. This shows that PhaseMO with \(V=1\) can model an analog beamformer.
    \item For \textbf{V=R}, we consider \(V\) as the number of physical RF chains of the hybrid beamforming architecture, thus the final equation can be written as \(\hat{Y}_{N\times 1}(f) = \Phi_{N\times R} \Gamma_{R\times K}(f) \mathbf{X}_{K \times 1}(f)\). This equation aligns with the form we have for a hybrid beamformer. It shows that PhaseMO with \(V=R\) can model a hybrid beamformer.
\end{itemize}

Notably, if we consider the channel, we can figure out what signal will be received on the user side:

\begin{equation}
    \hat{X}_{K\times 1} = H_{K \times N}(f) \Phi_{N\times V} \Gamma_{V\times K}(f) \mathbf{X}_{K \times 1}(f)
\end{equation}

Although the mathematical description assumes \name in a downlink scenario, by considering an ADC instead of a DAC and performing the desired interleaving, similar expressions can be derived for an uplink scenario as well. In conclusion, PhaseMO, with just a single RF chain, can adapt itself to different network throughputs in a softwarized manner.

\subsection{\textbf{Eliminating DAC non-idealities and FPS spreading effect}}
So far, we have shown the design of the \name and explained how using the fast phase changes can provide an adaptive MIMO architecture under the assumption that the spreading effect and DAC non-idealities can be ignored using a bandpass filter. Here, we demonstrate the effect of the DAC on the signal's spectrum, which represents a major issue that needs to be evaluated. This is particularly important because we use time-variant phase shifters, which can also introduce additional sidebands, more importantly, we want to make sure the BPF can eliminate the nonlinearity effects sourced by DAC non-ideality (i.e. function $A$) and FPSs. As later on, we have to consider the effect of a bandpass filter on the radiated signal, it is more convenient to do the mathematical analysis of this part in the frequency domain.

We already showed the frequency representation of the DAC output in Eq. \ref{eq: dac output frequency}.To add the DAC non-idealities to this expression, we can write the DAC output spectrum considering function \(A\)  which models the non-idealities coming from the DAC as \( \hat{Z}(f) = A(Z(f))\) which \(\hat{Z}(f)\) is the DAC output signal in the frequency domain.

In the next step, we will consider the FPS effect on the DAC output and observe how the spectrum is affected by the high-speed switching property of FPS. First, we express the time-domain signal at the output of the FPS and derive its frequency-domain representation as \(y_n(t) = \hat{z}(t) \times f_n(t) \xrightarrow{\mathscr{F}} Y_n(f) = \hat{Z}(f) * F_n(f)\) which $f_n(t)$ is the time-domain signal created by the FPS. As mentioned previously, each FPS toggles periodically between $V$ phases each of which with \(\frac{1}{Vf_s}\) duration; therefore, we can write down the FPS signal for the nth antenna in one period as follows: 

\begin{equation}
    f_n(t) = \sum_{v=0}^{V-1} e^{j\Phi_{nv}} \Pi_{T'_s}(t - vT'_s - \frac{T'_s}{2})
    \label{eq: FPS one period}
\end{equation}

Which \(f_n(t)\) is the time-variant signal produced by the FPS at the \(n\)-th antenna, \(\Phi_{nv}\) is the \(v\)-th phase produced by the \(n\)-th antenna FPS, and \(\Pi_{T'_s}\) demonstrates a pulse with width of \(T'_s\). Here, \(VT'_s\) is one-period duration, and \(f'_s=Vf_s\) is the DAC sampling frequency of PhaseMO with respect to conventional beamforming sampling frequency(\(f_s\)). 

Next, we can extend the Eq. \ref{eq: FPS one period} to account for the periodic nature of FPSs. In this regard, we can model the signal using the convolution of a single period with an impulse train which makes the further analysis of the process in the frequency domain easier. 

\begin{multline}
    f_n(t) = \sum_{i=-\infty}^{\infty} \sum_{v=0}^{V-1} e^{j\Phi_{nv}} \Pi_{T'_s}(t - iVT'_s - vT'_s - \frac{T'_s}{2}) \\
    = \sum_{v=0}^{V-1} e^{j\Phi_{nv}} \sum_{i=-\infty}^{\infty} \left(\Pi_{T'_s}(t - vT'_s - \frac{T'_s}{2}) * \delta(t - iVT'_s)\right)
    \end{multline}

Finallly, the frequency-domain representation of \(f_n(t)\) can then be derived as:

    \begin{multline}
    \resizebox{.95\hsize}{!}{$F_n(f) = \sum_{v=0}^{V-1} e^{j\Phi_{nv}} \sum_{i=-\infty}^{\infty} \mathscr{F}\left[\delta(t - iVT'_s)\right] \mathscr{F}\left[\Pi_{T_s}(t - vT'_s - \frac{T'_s}{2})\right]$} \\
    \resizebox{.95\hsize}{!}{$= \sum_{v=0}^{V-1} e^{j\Phi_{nv}} \sum_{i=-\infty}^{\infty} \frac{1}{VT'_s} \delta\left(f - \frac{i}{VT'_s}\right) T'_s e^{-j 2\pi f (vT'_s + \frac{T'_s}{2})} \text{sinc}(fT'_s)$} \\
    \resizebox{.95\hsize}{!}{$= \sum_{v=0}^{V-1} \frac{1}{V} e^{j\Phi_{nv}} \sum_{i=-\infty}^{\infty} \delta\left(f - \frac{i}{VT'_s}\right) e^{-j \frac{2\pi i}{VT'_s} (vT'_s + \frac{T'_s}{2})} \text{sinc}\left(\frac{i}{V}\right)$}
    \label{eq: fps created signal}
    \end{multline}

Substituting \(\hat{Z}(f)\) and \(F_n(f)\) into the \(Y_n(f) = \hat{Z}(f) * F_n(f)\) expression results in:

\begin{multline}
    Y_n(f) = \sum_{v=0}^{V-1} \frac{1}{V} e^{j\Phi_{nv}} \\
    \times \sum_{i=-\infty}^{\infty} A(Z(f))\bigg|_{f=f-\frac{i}{VT'_s}} \\
    \times e^{-j \frac{2\pi i}{VT'_s}} \text{sinc}\left(\frac{i}{V}\right)
\end{multline}

This expression indicates that the radiated signal's spectrum, which will be passed through the bandpass filter, includes the DAC's output spectrum and a \(\frac{i}{T'_sV}\) shifted versions of that, resulting from the time-variant phase shifters effect. The bandpass filter will remove all the sidebands out of \(B\) bandwidth centered at \(f_c\), so we need to determine the exact output of the filter.

Simplifying the result, we can observe that the DAC image artifacts are located at \(if'_s\) frequencies. For values of \(i\) that are multiples of \(V\) (\(i = \ldots, -2V, -V, 0, V, 2V, \ldots\)), the images will be shifted into the pass band of the filter. On the other hand, for these values of $i$, the Sinc function is only nonzero at $i!=0$ where we have \(A(Z(f)) \approx Z(f)\). Consequently, the output of the bandpass filter shows that the FPS switching combined with DAC non-idealities can be eliminated using the bandpass filter shown in Eq. \ref{eq: radiated signal wo nonideal} and used to prove the adaptability feature of \name.

\section{Evaluation}
\label{sec:evaluation}
So far, we have explained the design of the new beamforming architecture, \name, its operation, and its mathematical representation. In this section, we will discuss various simulation experiments conducted in the downlink scenario assuming perfect channel estimation feedback from the user side to verify \name's design and demonstrate its key application: load-adaptable power consumption while maintaining good throughput. First, we will describe the evaluation setting. Then, we will compare \name's throughput with various baselines, including digital beamforming (DBF), fully connected hybrid beamforming (HBF), partially connected hybrid beamforming, analog beamforming (ABF), and GreenMO \cite{gupta2023greenmo}. We will also examine how \name achieves reasonable energy efficiency by adapting to network traffic while maintaining coverage area and UE transmit power consumption. Additionally, we will compare \name's adaptability with antenna muting combined with digital beamforming as the baseline.

\subsubsection{\textbf{Evaluation setting}} 
\begin{figure}
    \centering
    \includegraphics[width=\linewidth]{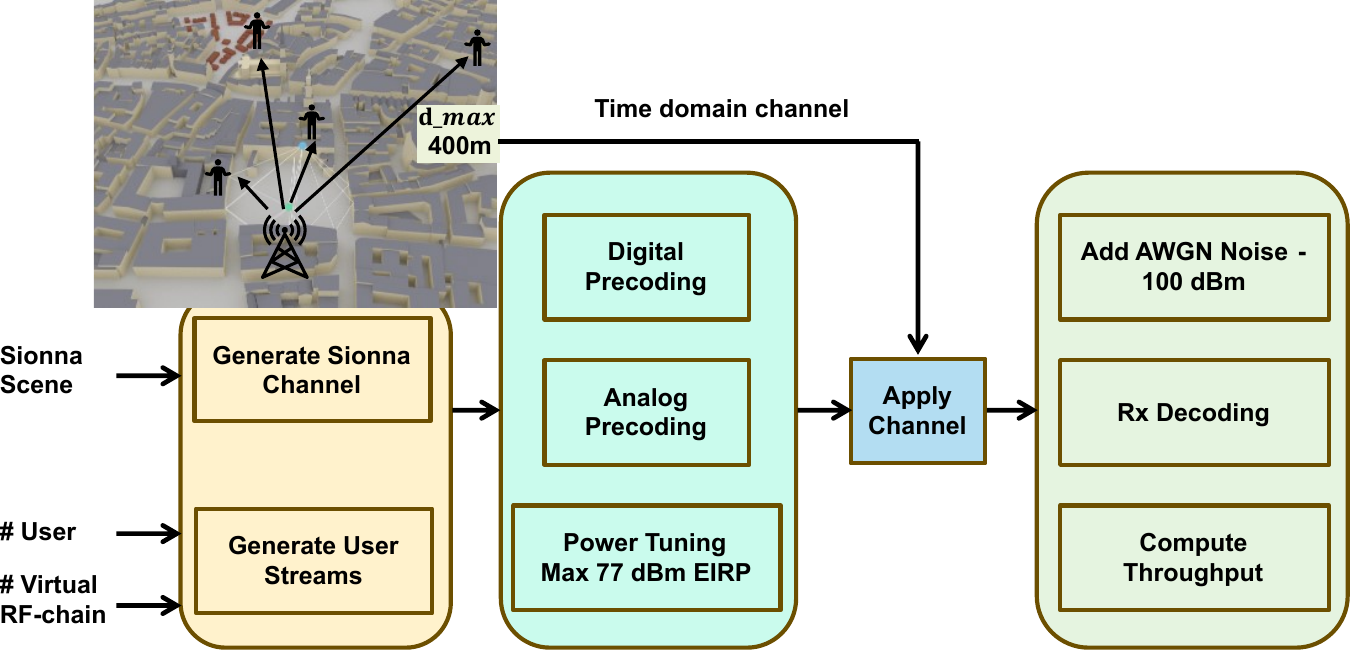}
    \caption{ Evaluation setup using Sionna channel in MATLAB: Users' data is precoded into OFDM symbols (100 MHz), passed through phase shifters, amplified, and transmitted. SINR at users is converted to spectral efficiency and throughput via the 5G-NR MCS table.
    }
    \label{fig:eval setup}
\end{figure}

To evaluate \name, we utilize the frequency channel derived from Sionna, a GPU-accelerated open-source library for link-level simulations based on TensorFlow. Sionna generates the wideband channel frequency response in an open environment model that includes buildings of various sizes. In this study, we set up a base station at a height of \(35\)~meters, equipped with \(64\) antennas, operating at 4.2~GHz, and consider the distribution of single-antenna users at varying distances from the base station in the provided Munich city map. Specifically, we place one user randomly at a certain distance \(d\) from the base station, ensuring that the remaining users are located within a distance less than \(d\). Sionna identifies all beams that can reach the users from the base station and determines the channel impulse response, which includes multiple taps with different phases and attenuations. Finally, it obtains the channel information for each subcarrier based on parameters such as the center frequency (4.2~GHz), number of subcarriers (64), and subcarrier spacing. This configuration is repeated 10 times for each \(d\), as shown in Fig.~\ref{fig:eval setup} for the case of 4 users. The evaluation setup, as illustrated in Fig.~\ref{fig:eval setup}, allows us to comprehensively evaluate \name in a realistic over-the-air wireless channel scenario.

As shown in Fig. \ref{fig:eval setup}, we use Sionna channel frequency response in the MATLAB simulation platform to implement different beamforming techniques. Initially, the random users' bits shaped in 64-QAM constellation points are precoded digitally in orthogonal frequency division multiplexing (OFDM) symbols, each with a \(100\) MHz bandwidth. Then, the OFDM symbols, after conversion to the time domain, pass through the analog phase-shifter network (i.e., analog or hybrid beamforming). Finally, the filtered radiated signal from each antenna is amplified for over-the-air transmission conditioned not violating maxEIRP of \(77\)dBm limit \cite{ntia2022}.

On the user side, additive white Gaussian noise (AWGN) with \(-100\)~dBm power is added to the received signals. The Error Vector Magnitude (EVM) of the received 64-QAM constellation is computed to derive the user's Signal to Interference + Noise Ratio (SINR)~\cite{brown2018prediction}. The SINR is then mapped to spectral efficiency and net throughput using the 5G-NR modulation and coding scheme (MCS) table \cite{thyagarajan2021snr}.

\subsubsection{\textbf{Throughput evaluation}}Here, we evaluate \name in terms of net throughput alongside other beamforming techniques, including DBF, fully connected HBF, partially connected HBF, ABF, and GreenMO. As discussed earlier, to calculate the spectral efficiency (SE) in bps/Hz for each user, we use the 5G-NR MCS table to convert SINR to SE for each user. We then sum the spectral efficiency of all users and consider the total bandwidth used to obtain the net throughput of the system. This process is repeated for different distances from the base station and is averaged over 10 different user configurations.

For each beamforming technique, the corresponding approach is applied. For all beamforming methods, we use the same Zero Forcing (ZF) equalization for digital precoding based on the channel per subcarrier derived from Sionna, while considering the impact of analog precoding. However, we do not utilize intelligent techniques in the analog precoder and use ad-hoc approaches for all methods. For example in GreenMO, since we do not have the BABF approach used in the original design, we instead use a random invertible precoding binary matrix, which is supposed to achieve performance relatively close to the original work. For ABF, hybrid beamforming, and PhaseMO, we employ the center subcarrier channel phase as the analog beamforming approach. For partially connected HBF, we use a combination of the ABF and PhaseMO approaches to generate the analog beamformer design. This ensures a consistent setup for comparisons across all tested methods.

\begin{figure}
    \centering
     \includegraphics[width=0.8\linewidth]{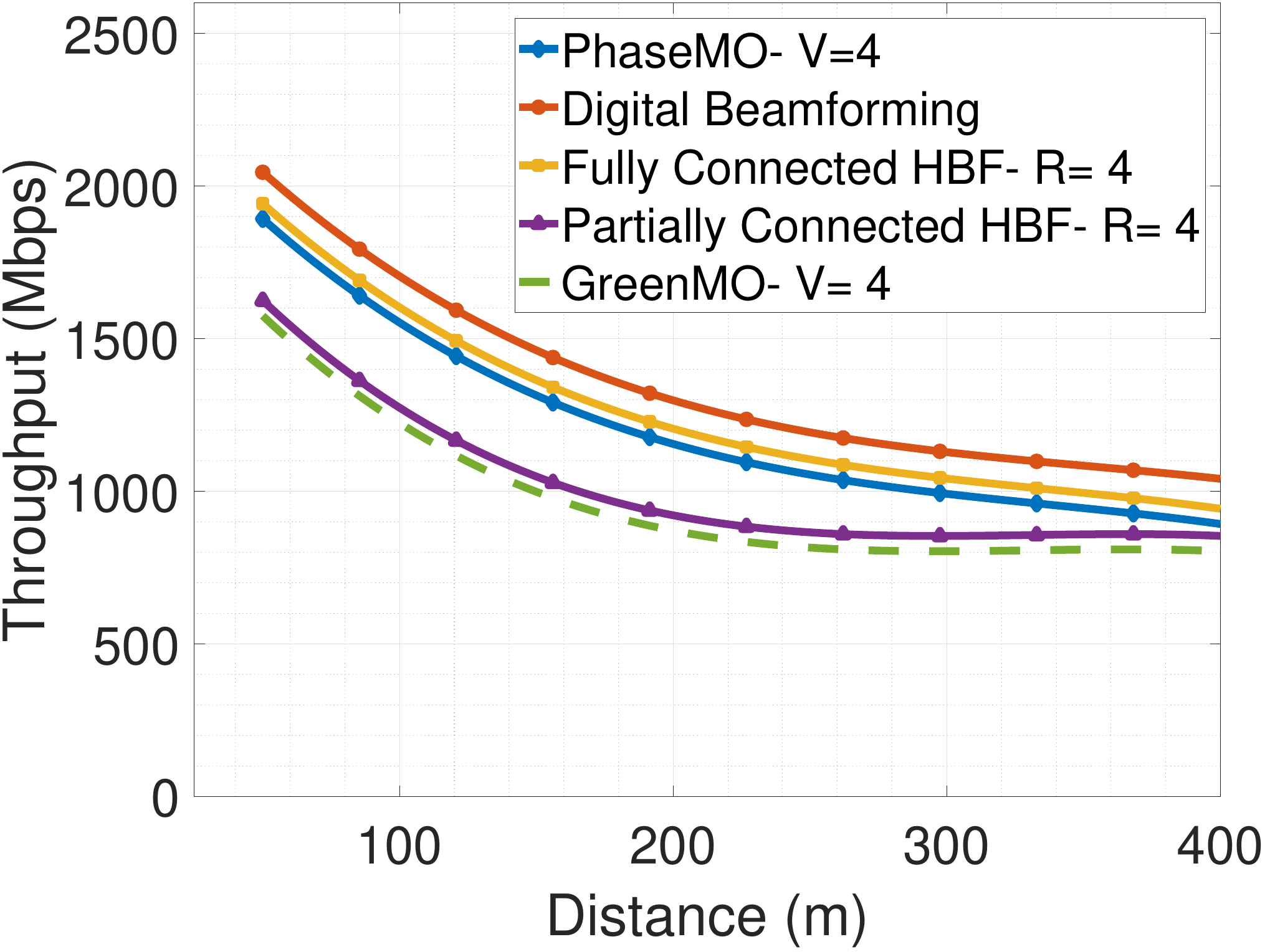}
    \caption{Net throughput for different beamformings in fixed number of users(\(4\)) network, which shows \name can achieve FCHBF throughout. Figure includes the curves for different beamformings considering 64 antennas and physical or virtual RF chains (\(V<N\)) which shows that FCHB and \name with relatively similar throughput have poorer performance compared to DBF.
    }
    \label{fig: BFcompare}
\end{figure}

In Fig. \ref{fig: BFcompare}, \name's net throughput which employs \(V=4\) virtual RF chains with 64 antennas compared to other baselines for \(4\) users is shown. It is worth noting that, we considered \(4\) physical RF chains for hybrid beamforming methods in this evaluation. It can be observed that the \name net throughput is a little bit lower than fully-connected HBF, also it can be observed that partially connected HBF and GreenMO with approximately the same throughput stand below \name as mentioned previously in \cite{gupta2023greenmo}. Due to the more strict requirement of using filters in \name, we account for SINR degradation due to bandpass filter insertion loss considering the 40-45 dB adjacent channel power ratio (ACPR) attenuation for C-band communications~\cite{ntia2022}.

\subsubsection{\textbf{Adaptability evaluation}} To evaluate the adaptability feature of \name, we analyze throughput degradation and operational power decrease while reducing the number of virtual RF chains. On the other hand, we evaluate the same metrics while using antenna muting combined with digital beamforming (AM +DBF). Finally, combining the power and throughput evaluation, we show how reducing the number of physical RF chains in AM +DBF or virtual RF chains in \name will affect the energy efficiency (EE) in different load scenarios. In addition, we show how coverage area and UE transmit power will be affected in AM + DBF.

\begin{figure}
    \centering
     \includegraphics[width=\linewidth]{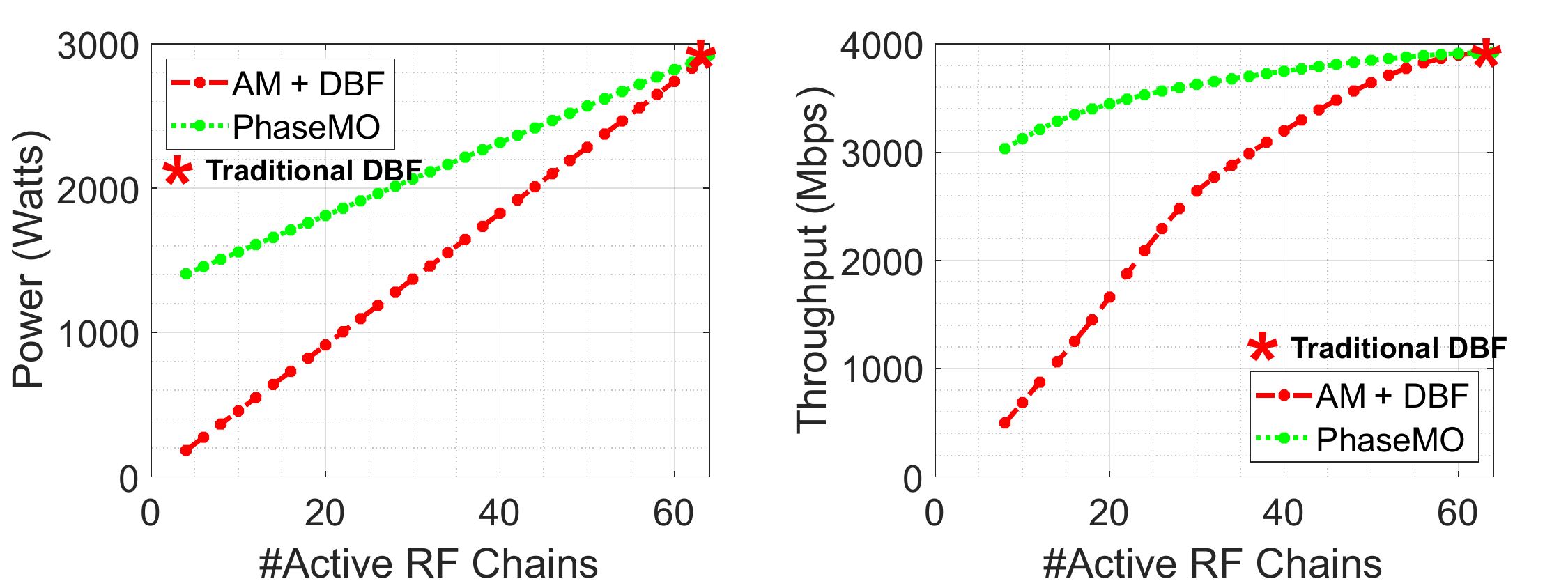}
    \caption{a) Power consumption, and b) Throughput of antenna muting combined with digital beamforming and \name versus different numbers of unmuted/virtual RF chains for 8 users is depicted which shows \name can maintain a good throughput while reducing the power; however, AM+DBF reduces more power at the cost of throughput reduction by a lot}
    \label{fig: power}
\end{figure}

Figure \ref{fig: power}.a) shows the power consumption of a traditional 64-antenna BS for varying numbers of active/virtual RF chains. For a base station, the total power consumption is mostly dominated by PAs and base-band processing power consumption. PA power consumption can be calculated based on the number of active RF chains, PA's output power which is defined based on the max EIRP limit, and PA's efficiency. Here, we consider \(60\)\% power efficient PAs (\cite{kim2020high}) and put the output power of each PA based on maxEIRP of \(77\)dBm. On the other hand, base-band processing power consumption can be evaluated using \cite{zorello2022power} and \cite{bjornson2016massive}. We consider \(15\)GFLOPS for each number of active/virtual RF chains each of which requires \(1.683\)W. So, we can compute the BS power consumption taking into account the power consumption of  PAs' and base-band processing. 

Although \name can save about 5-7\% power compared to full-capability DBF (all antennas are active), we get more power efficiency comparing its adaptability property with AM+DBF. Taking advantage of AM+DBF to mitigate the power consumption, as we reduce the number of active physical RF chains, both PA power consumption and base-band processing power will drop; however, due to reducing the number of active PAs in the architecture, and multiplexing gain, the system's net throughput will be reduced as well (Fig. \ref{fig: power}.b). On the other hand, in \name similar to hybrid beamforming, if we reduce the number of virtual RF chains, we gain only from reducing the base-band processing power consumption while maintaining all the PAs active which won't cause any throughput reduction. 

\begin{figure}
    \centering
     \includegraphics[width=\linewidth]{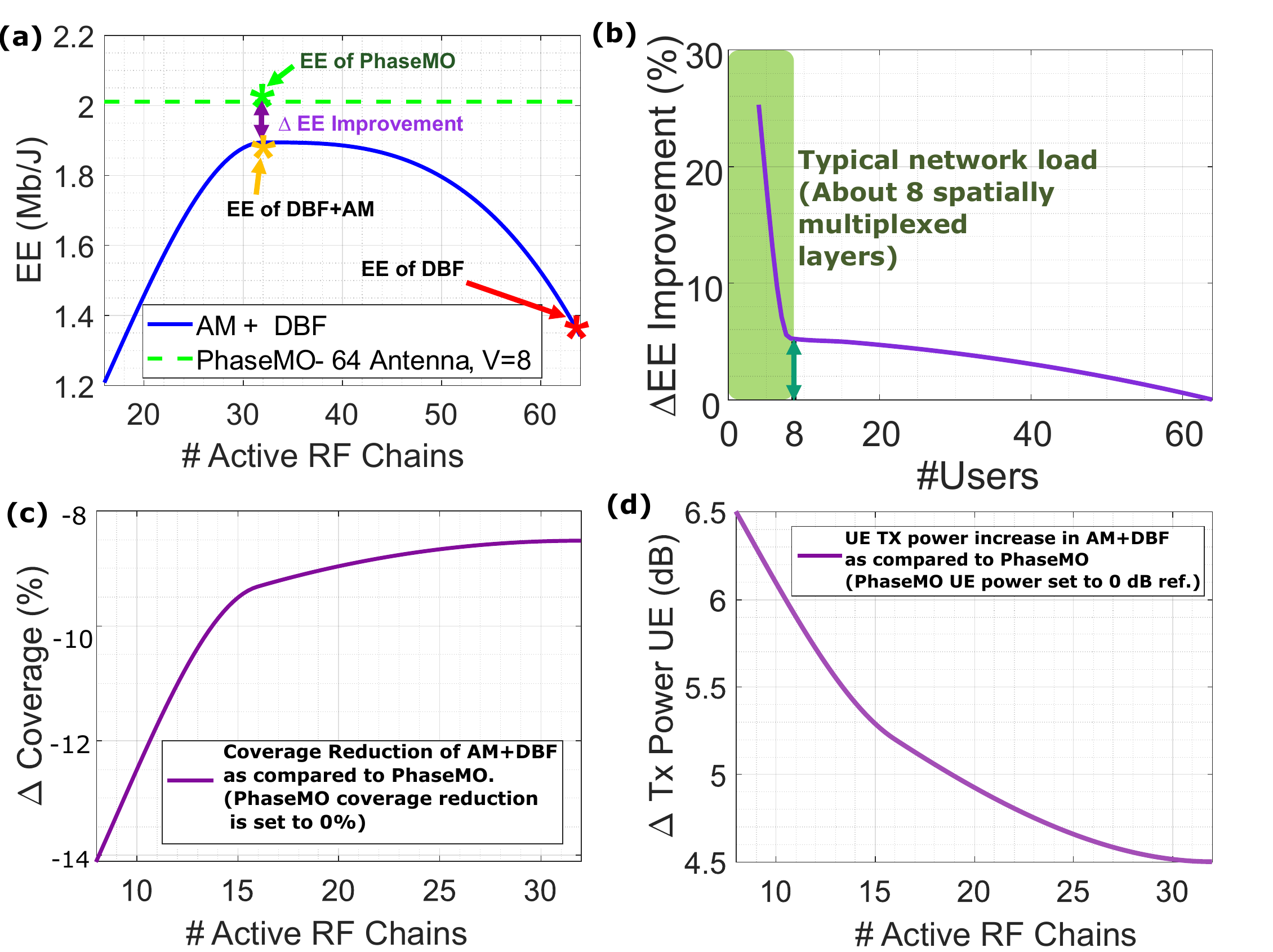}
    \caption{a) \name shows better energy efficiency (EE) assuming an equal number of users and virtual RF chains, even when DBF is optimized using AM for 8 users, b) \name can achieve at least \(5\)\% EE improvement for up to 8 users with respect to AM+DBF best point of operation, c) AM+DBF will lose at least \(8\)\% coverage area by turning off the half number of RF chains due to less number of active PAs comparing to 64 RF chains full capacity point of operation, d) AM+DBF will cause at least \(5\)dB power increase in UE transmit power consumption due to losing multiplexing gain by reducing half of the BS antennas comparing to 64 RF chains full capacity point of operation}
    \label{fig:AM effect}
\end{figure}

Therefore, we can evaluate which technique works better in terms of saving power while keeping a reasonable throughput, thus we use energy efficiency (\(\frac{b}{J}\)) which combines both of these parameters. As shown in Fig. \ref{fig:AM effect}.a) for a network with \(8\) number of users, using \name with constant \(8\) virtual RF chains maintain an energy efficiency higher than the optimum operation point of AM+DBF which is when half of the BS RF chains (\(32\) out of \(64\)) are muted, Consequently, as shown this result in Fig. \ref{fig:AM effect}.b) for different number of users, as we increase the number of users up to \(8\) which is typical number of users connected to today's Massive MIMO BSs, due to less throughput gain of \name with respect to AM+DBF, the EE improvement gain for \name in comparison with the best operating point for AM+DBF will reduce to 5\% which still is promising.

Although 5\% does not look like a promising improvement if we replace AM+DBF with \name for a case of 8 users, \name can be considered a better adaptive solution if we include other performance metrics such as coverage area and UE transmit power in uplink. As shown in Fig. \ref{fig: power}.b), AM+DBF loses more throughput while trying to achieve a low power consumption due to turning off a few of the antennas and reducing the multiplexity gain. Therefore, it also reduces the coverage area on one side, and on the other side, it makes UE devices draw more power to achieve the same throughput in uplink; however, since \name just loses some throughput due to less number of virtual RF chains, it won't affect the coverage area and UE transmit power. 

As shown in Fig. \ref{fig: power}.c), if we consider the base-station (BS) coverage map with 64 antennas as the baseline in Sionna, due to reducing the number of RF chains the coverage area will decrease and it even reaches about 14\% coverage reduction. Additionally, considering the same baseline if we randomly distribute users in the base-station coverage area, the average transmit power of the user will increase by 4.5dB in the best case and 6.5dB in the worst case.  As a result, \name guarantees the worst-case coverage and throughput requirements while improving the energy efficiency of the network. This is achieved by designing an architecture that always uses a maximum number of antennas while flexibly cutting down on digital processing power.
    
\section{Discussion and Limitations}
\label{sec:discussion}

In this section, we examine the implications of the \name architecture, acknowledging its limitations and providing insights for future research and potential enhancements.

\begin{itemize}
    
    \item \textbf{Scaling to Higher Number of Antennas:} While \name leverages a single RF chain, scaling beyond 64 antennas is hindered by hardware and design limitations, including the high DAC sampling frequency requirement, phase shifter bandwidth, and power spreading effects. The maximum available DAC sampling frequency today is 6.4~Gsps~\cite{analog2024ad9162}, which supports up to 64 antennas with \(100\)~MHz over-the-air bandwidth. Similarly, the FPS modulation bandwidth, currently limited to 2.5~GHz~\cite{hmc877lc3}, restricts interfacing to 24 antennas using the current circuit design. Moreover, FPS-induced power spreading attenuates the main band signal's power by \(\frac{1}{V^2}\), requiring small pre-amplifiers combined with a larger PA providing 47~dB gain. Under these conditions, supporting more than 128 users becomes challenging. To overcome these limitations, a hybrid architecture that combines PhaseMO with sets of antennas, each supported by a separate RF chain, can be adapted to enable scalability to higher numbers of antennas.

    \item \textbf{Handling Out-of-Band Emission:} Due to the spectral spreading effect of FPSs, high-precision narrowband (\(100\)~MHz) bandpass filters with high out-of-band attenuation are necessary to reduce ACPR. However, the non-ideal characteristics of these filters can worsen interference. A potential solution is to use a higher sampling frequency instead of \(V \times\) the signal bandwidth, which simplifies the filtering process and mitigates this limitation.

    \item \textbf{Channel Estimation compliance with \name:} The 5G standard for MIMO-based channel estimation at present assumes one antenna mapped to one digital port for channel estimation, which makes it difficult to integrate architectures like \name, and as well hybrid beamformers. However, there are new changes proposed in release 18 and beyond \cite{jin2023massive}, like orthogonal cover codes can enable more antenna channels from fewer ports, which can make \name compliant with 5G.

    \item \textbf{Comparison Scope and Power-Saving Techniques:} This work focuses on throughput comparisons with a few selected methods and evaluates power-saving techniques primarily against antenna muting. Furthermore, the precoding techniques used in this study are not chosen optimally. Extending the comparisons to include approaches such as power backoff, other advanced power-efficient methods, and intelligent precoding schemes would provide a more comprehensive analysis and further validate the architecture's performance.
\end{itemize}

\section{Conclusion}

This work introduces \name, a novel Massive MIMO architecture that dynamically optimizes power consumption based on network load without compromising key performance metrics such as coverage, throughput, and user-device power. By leveraging Fast Phase Shifters (FPS), \name ensures full array beamforming gain while minimizing digital interfacing in line with network load conditions. Our results demonstrate that \name achieves an energy efficiency improvement of up to 30 \% in low-load scenarios while avoiding approximately 10 \% coverage reduction and a 5 dB increase in UE transmit power. Additionally, \name addresses critical limitations of existing approaches, such as antenna muting and hybrid beamforming, by overcoming hardware scalability challenges and ensuring adaptability to rapidly evolving network demands. These advancements position \name as a practical and scalable solution for achieving energy-efficient and high-performance cellular networks in the era of increasing connectivity needs.

\section*{Acknowledgment}

This work was supported in part by the U.S. National Science Foundation (NSF) under Award No. 2211805. Further, the authors acknowledge the anonynmous reviewers and WCSNG group members for their feedback.

\printbibliography

\end{document}